\documentclass[aps,pre,twocolumn,superscriptaddress]{revtex4}
\usepackage[utf8]{inputenc}
\usepackage{amsmath}
\usepackage{amssymb}
\usepackage{mathtools}
\usepackage{graphicx}
\usepackage{cases}
\usepackage{color}

\usepackage{hyperref}

\DeclareMathOperator{\tr}{tr}
\DeclareMathOperator{\sgn}{sgn}

\begin{document}

\title{Cooper pair splitting in diffusive magnetic SQUIDs}

\author{P.\ A.\ Ioselevich}
\affiliation{National Research University Higher School of Economics,
Myasnitskaya str.\ 20, Moscow, 101000 Russia}
\affiliation{L.\ D.\ Landau Institute for Theoretical Physics, Kosygin str.\ 2,
Moscow, 119334 Russia}

\author{P.\ M.\ Ostrovsky}
\affiliation{Max-Planck-Institute for Solid State Research, Heisenbergstr.\ 1,
Stuttgart, 70569 Germany}
\affiliation{L.\ D.\ Landau Institute for Theoretical Physics, Kosygin str.\ 2,
Moscow, 119334 Russia}

\author{Ya.\ V.\ Fominov}
\affiliation{L.\ D.\ Landau Institute for Theoretical Physics, Kosygin str.\ 2,
Moscow, 119334 Russia}
\affiliation{Moscow Institute of Physics and Technology, Institutsky per.\ 9,
Dolgoprudny, 141700 Russia}

\author{M.\ V.\ Feigel'man}
\affiliation{L.\ D.\ Landau Institute for Theoretical Physics, Kosygin str.\ 2,
Moscow, 119334 Russia}
\affiliation{Moscow Institute of Physics and Technology, Institutsky per.\ 9,
Dolgoprudny, 141700 Russia}

\date{2 April 2018}

\begin{abstract}
We study Josephson junctions with weak links consisting of two parallel disordered arms with magnetic properties -- ferromagnetic, half-metallic or normal with magnetic impurities. In the case of long links, the Josephson effect is dominated by mesoscopic fluctuations. In this regime, the system realises a $\varphi_0$ junction with sample-specific $\varphi_0$ and critical current. Cooper pair splitting between the two arms plays a major role and leads to $2\Phi_0$ periodicity of the current as a function of flux between the arms. We calculate the current and its flux and polarization dependence for the three types of magnetic links.
\end{abstract}

\maketitle
\section{Introduction}
Superconductivity implies natural entanglement and coherence between electrons in Cooper pairs. This opens up possibility for realization of EPR pairs \cite{EPR} and testing Bell inequalities \cite{Bell} in solid-state devices. The most spectacular effects are expected when Cooper pairs split, so that electrons are spatially separated between two different ``arms'' (e.g., in forklike geometry). The splitting can be enforced by energy filtering or spin filtering inside the arms [by means of electrically tunable quantum dots or ferromagnetic filters, respectively]. Peculiarities of correlated transport through the arms of such multiterminal devices have been studied both theoretically \cite{Recher2001,Lesovik2001,Borlin2002,Chtchelkatchev2002} and experimentally \cite{Hofstetter2009,Herrmann2010,Schindele2012,Das2012}. Josephson current through two arms containing quantum dots (where the Coulomb energy impedes passage of nonsplit Cooper pairs through each arm) was theoretically studied in Refs.~\cite{SplitterQD}.

While the main attention up to now has been paid to the quantum-dot scheme of Cooper pair splitting, the ferromagnetic realization \cite{Lesovik2001,Chtchelkatchev2002} has certain advantages. Half-metallic (H) ferromagnets (already employed in various superconducting hybrid structures, see, e.g., Refs.\ \cite{Keizer2006,Singh2015}) should lead to highly efficient splitting due to absolute spin filtering. At the same time, mutual orientation of magnetizations in the two arms can, in principle, be varied continuously by weak external magnetic field if one of the arms is exchange biased (a scheme similar to, e.g., Ref.\ \cite{Leksin2012}). This provides an additional degree of freedom for controlling the device.

The dependence of Cooper pair splitting on the magnetization orientation of the arms has been studied experimentally in an SF setup \cite{BWL2004,BL2005}. Two ferromagnetic arms F were contacted to a superconductor S close to each other and a voltage was applied to one of the arms. This produced a current in the other ferromagnetic arm due to crossed Andreev reflection. The current was sensitive to the relative orientation of the magnetizations.

In this paper, we take this setup a step further and consider SQUID geometry, i.e., a Josephson junction with two diffusive magnetic arms $a$ and $b$ connecting the superconducting leads, as schematically depicted in Fig.~\ref{fig1}. We consider three versions of the SQUID, with arms made of ferromagnet (F), half-metallic ferromagnet (H), or normal metal with magnetic impurities (M). In all the three cases, the disorder-averaged Josephson current is suppressed. At the same time, any particular sample exhibits a current originating from mesoscopic fluctuations. The aim of this paper is to calculate this current and study its dependence on phase difference $\varphi$, on magnetization orientation (for the F and H systems) and on the magnetic flux $\Phi$ threading the SQUID.

We are interested in properties of coherent supercurrent transport through the whole system. At the same time, additional information, related to entanglement between electrons from the same Cooper pair, could be extracted from the correlations between the currents in the two magnetic arms. However, these correlations and hence electron entanglement are beyond the scope of this paper.

The supercurrent in a generic Josephson junction is carried by Cooper pairs travelling across the junction. The current is comprised of contributions from different Cooper-pair trajectories. In a conventional SNS junction (where N stands for normal metal without magnetic impurities), all contributions come with the same sign due to time reversal symmetry (TRS).

If TRS is broken, contributions from different Cooper pairs come with random phases, and the total current is suppressed. Indeed, for an SMS or single-domain SFS junction the disorder-averaged current decays exponentially with length, $\langle I\rangle\propto e^{-L/l_s}$. Here $l_s$ is the diffusive spin-flip scattering length in the M case, and $l_s=\sqrt{\hbar D/h_\mathrm{ex}}$ in the F case, where $D$ is the diffusion constant and $h_\mathrm{ex}$ is the exchange field. However, while $\langle I\rangle$ in a long SFS junction is exponentially small, the typical current in a specific sample is much larger and does not contain exponential smallness \cite{Zyuzin2003}.

A half-metal is fully polarized, therefore an SHS junction cannot conduct $s$-wave Cooper pairs. This brings us to the SQUID geometry, Fig.~\ref{fig1}, which allows supercurrent to flow via pair splitting
(physically equivalent to the crossed Andreev reflection), provided the arms are made of differently polarized half-metals. The disorder-averaged current in a similar system has been previously found to be suppressed very strongly, $\langle I\rangle \propto e^{-L/l}$, where $l$ is the mean free path \cite{Melin2003}. However, as we will show below, the sample-specific current contains no such smallness.

The paper is organized as follows. In Sec.\ II, we formulate our method for calculating supercurrents in the SQUID. In Sec.\ III, we present our results. In Sec.\ IV we discuss the details of the obtained results and their experimental implications. Finally, we present our conclusions in Sec.\ V.

\section{Method}

To study the typical current, we are going to calculate the correlator $\langle I(\varphi_1,\Phi_1)I(\varphi_2,\Phi_2)\rangle$ between total currents (including both arms of the SQUID) taken at different phases $\varphi_1,\varphi_2$ and fluxes $\Phi_1,\Phi_2$. This correlator (which we will denote $\langle I_1I_2\rangle$) captures the mesoscopic fluctuations of the current and will be used to restore the shape and amplitude of $I(\varphi,\Phi)$ in the SQUID. In what follows, exponentially small values are neglected, e.g., we write $\langle I\rangle=0$.

\begin{figure}
\centering
\hspace*{-3pt}\includegraphics[width=0.485\textwidth]{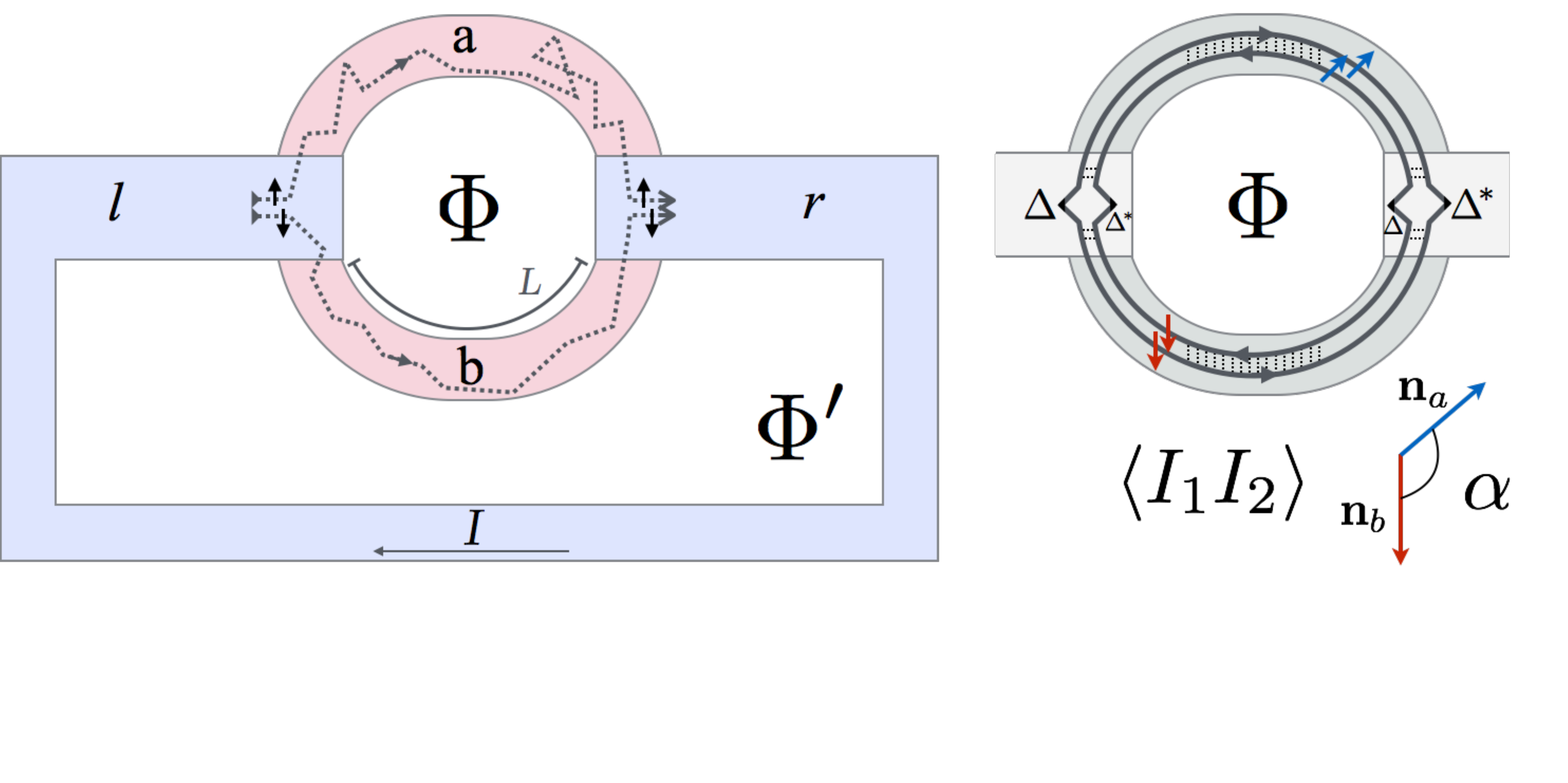}
\caption{
Left: Schematic of the system. Two magnetic arms $a$ and $b$ (red) form a Josephson junction between superconductors $l$ and $r$ (blue).
We study the supercurrent $I(\varphi,\Phi)$, with the superconducting phase difference $\varphi$ controlled by the flux $\Phi'$.
In the case of half-metallic arms, the supercurrent is carried by Cooper pairs that split between the arms. Right: Diagram representing the current-current correlator in the half-metal case. It describes split-pair transport and consists of two polarized diffusons travelling through the arms, joined in the superconducting leads by anomalous Green's functions.
}
\label{fig1}
\end{figure}

Quasiclassical methods allow studying average values, such as $\langle I(\varphi)\rangle$. To calculate $\langle I_1I_2\rangle$, a more involved method is required. We use the nonlinear $\sigma$ model here. The supercurrent in a system is related to its free energy via $I(\varphi)=(2e/\hbar)\partial F/\partial\varphi$. Using this relation, we write the current-current correlator using the replica $\sigma$ model \cite{Wegnersigma}:
\begin{align}
\langle I_1I_2\rangle = \frac{e^2}{\hbar^2}T^2\sum\limits_{\omega_1,\omega_2}\frac{\partial^2 }{\partial\varphi_1\partial\varphi_2}\lim\limits_{n_{1,2}\to0}\frac{\int e^{-S[Q]}DQ}{n_1n_2}. \label{IQ}
\end{align}
The supercurrents $I_1$ and $I_2$ are the results of summation over the Matsubara energies $\omega_1$ and $\omega_2$, respectively. The field $Q(x)$ is a matrix in replica space containing $n_1$ replicas corresponding to $I_1$ and another $n_2$ replicas corresponding to $I_2$. There are also spin space (Pauli matrices $s_i$), Nambu space ($\tau_i$), and particle-hole space ($\sigma_i$) taking into account the Bogoliubov--de Gennes symmetry of the problem \cite{Altlandsigma}. $Q$ further obeys the nonlinear condition $Q^2=1$ and the linear constraint $Q=CQ^TC^T$ with $C=i\tau_1\sigma_1s_2$. Temperature $T$ in Eq.\ (\ref{IQ}) is measured in energy units (i.e., $k_B =1$).

The action $S[Q]$ of the $\sigma$ model consists of bulk terms $S_i$ describing the two superconducting leads $l,r$ and the two magnetic arms $a,b$ of the SQUID, and boundary terms $S_{ij}$ describing the four superconductor/arm interfaces in the system:
\begin{align}
S =\sum S_i+\sum S_{ij}\label{actionQ}.
\end{align}
The superconducting leads are described by
\begin{align}
S_i = \frac{\pi\nu_s}{8}\int\tr\left[\hbar D_s(\nabla Q)^2-4(\hat{\omega}\sigma_3\tau_3+\check{\Delta}_i)Q\right]dx,
\end{align}
with $i=l,r$. Here $\nu_s$ is the one-dimensional normal-state density of states (it is related to the three-dimensional $\nu_{3d}$ as $\nu_s=\nu_{3d}A$ with $A$ being the cross-section area), $D_s$ is the diffusion constant, and $\check{\Delta}_{i} = |\Delta|(\tau_x\cos\hat\varphi_i-\tau_y\sin\hat\varphi_i)$. The Matsubara-energy operator $\hat\omega$ equals $\omega_1$ in the first $n_1$ replicas and $\omega_2$ in the other $n_2$ replicas. The same applies to the superconducting phases $\hat{\varphi}_i$ and the flux $\hat{\Phi}$.

The action in the magnetic arms reads
\begin{align}
S_j=\frac{\pi\nu}{8}\int\tr\left[\hbar D(\nabla Q)^2-4\hat{\omega}\sigma_3\tau_3Q\right]dx,\label{SF}
\end{align}
with $j=a,b$. The action \eqref{SF} is supplemented by the constraint $[Q,\tau_3]=0$. This owes to spin-flip scattering in the M case, and to the effect of orbital magnetic fields in the H and F cases. The magnetization $M$ required to justify the constraint for an arm of width $w\sim 100$\,nm and length $L\sim 1$\,$\mu$m is $M\sim \Phi_0/Lw \approx 20$\,mT with $\Phi_0=\pi\hbar c/e$ being the superconducting flux quantum. We expect typical magnetization in experiment to be larger than that.

In what follows we derive $\langle I_1I_2\rangle$ for the H and M cases. The F case is then deduced from the H-case results.

The difference between H and M arms lies in the spin structure of $Q$. In a half-metal arm, all conducting electrons are in the same spin state $|\mathbf{n}\rangle$, polarized along $\mathbf{n}$. In an M arm, magnetic impurities flip spins, which leads to a $Q$ matrix with trivial spin structure. As a result, the explicit spin structure in the two cases is
\begin{subnumcases}{\label{spinAll} Q=}
 |\mathbf{n}\rangle q \langle\mathbf{n}|, &H,\label{spinH}\\
 s_0q, &M,\label{spinM}
 \end{subnumcases}
where $q$ acts in spaces other than spin.

For simplicity, we assume that all four superconductor/arm interfaces are tunneling contacts with the same total conductance $G_t$, so that \cite{Efetov}
\begin{align}
S_{ij} = - \frac{g_t}{8}\tr\left[Q_i e^{\pi i\tau_3\frac{\hat\Phi_{ij}}{\Phi_0}} Q_je^{-\pi i\tau_3\frac{\hat\Phi_{ij}}{\Phi_0}}\right],
\end{align}
where $i=l,r$ and $j=a,b$, and the dimensionless conductance is $g_t=2G_t\pi\hbar/e^2$ for H and $g_t=G_t\pi\hbar/e^2$ for M. The two $Q$ matrices under the trace are taken on the two sides of the interface. The exponents account for the flux $\hat\Phi$. We use a gauge where the vector potential change is located at the contacts, $\hat\Phi_{la}=\hat\Phi_{rb}=\hat\Phi/2$ and $\hat\Phi_{lb}=\hat\Phi_{ra}=0$. In this gauge, the phase difference is
$\varphi = 2\pi\Phi'/\Phi_0+\pi\Phi/\Phi_0$. This choice is natural for studying pairs split between the two arms.

\begin{figure}
\centering
\hspace*{-3pt}\includegraphics[width=0.485\textwidth]{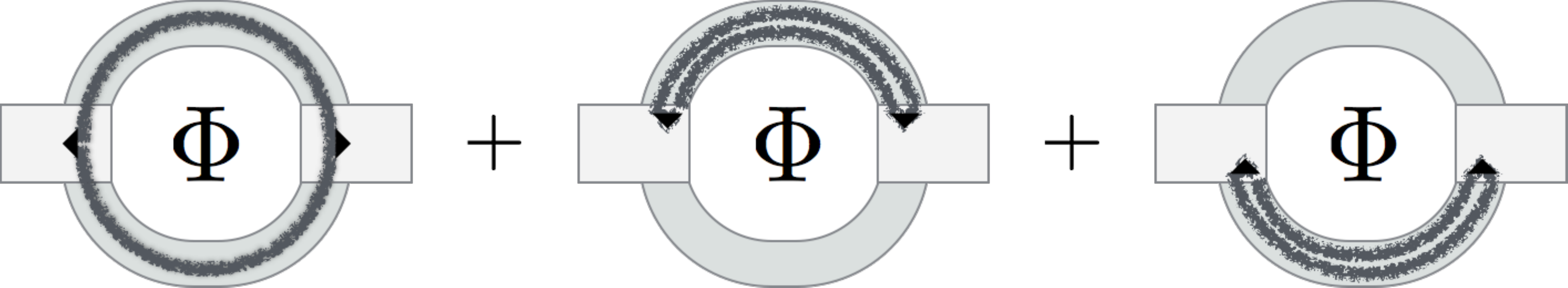}
\caption{
Diagram of the current-current correlator (\ref{IIM}) for arms with magnetic impurities. Thick lines represent diffusons. Leftmost diagram is the same as in Fig.~\ref{fig1} except the diffusons in M are singlet as opposed to polarized in H. The other two diagrams place both diffusons in the same arm and describe processes without pair splitting. The three diagrams give rise to the three terms in Eq.\ \eqref{IM}.
}
\label{fig2}
\end{figure}

To calculate the correlator \eqref{IQ}, we start with the saddle-point configuration of the bulk action $\sum S_i$. The solution is constant, $Q(x)=\mathcal{Q}$ in each of the four regions. In the superconductors,
\begin{align}
\mathcal{Q}_i = \sigma_3\tau_3\cos\hat\theta +
\sigma_0(\tau_1\cos\hat\varphi_i-\tau_2\sin\hat\varphi_i)\sin\hat\theta, \label{saddleS}
\end{align}
where $\hat\theta = \arctan(\hat\omega / |\Delta|)$. In the arms, the saddle-point solution $\mathcal{Q}$ is given by Eq.\ \eqref{spinAll} with $q=\sigma_3\tau_3\sgn\hat\omega$. The saddle-point approximation of the $\sigma$ model corresponds to the quasiclassical solution. In particular, the saddle-point equations coincide with the Usadel equations \cite{Usadel1970} with $\theta$ being the Usadel angle parameter. In our system, this approximation yields a zero supercurrent.

To go beyond quasiclassics, we need to integrate over fluctuations near the saddle point. We parameterize the fluctuations as $Q=e^{-iW/2}\mathcal{Q}e^{iW/2}$, where the fluctuation matrix $W(x)$ is much smaller than unity. We choose $W$ to anticommute with the saddle point $\mathcal{Q}$ (which is uniform in each of the four regions). In addition, $W$ obeys $W=-CW^TC^T$. In the magnetic arms, we also have $[W,\tau_3]=0$. Physically, this means that cooperons are suppressed by orbital magnetic fields or magnetic impurities. Cooperons are soft modes describing the diffusive propagation of electron pairs. They are responsible for the proximity effect in dirty SN junctions \cite{Altlandsigma}. In our magnetic arms, pairs break apart due to the exchange or orbital fields on the short length scale $l_s\ll L$, which allows us to neglect the exponentially weak effect of cooperons altogether, simply imposing $[Q,\tau_3]=0$ (and consequently $[W,\tau_3]=0$) in the magnetic arms.

Expanding $S$ near $\mathcal{Q}$, we get an effective action for $W$. The bulk action becomes
\begin{align}
S_i = \frac{\pi\nu_s}{8}\int \tr\left[\hbar D_s(\nabla W)^2+2\sqrt{\hat\omega^2+|\Delta|^2}W^2\right]dx\label{SWS}
\end{align}
in the superconductors and
\begin{align}
S_j = \frac{\pi\nu}{8}\int \tr\left[\hbar D(\nabla W)^2+2|\hat\omega|W^2\right]dx\label{SWF}
\end{align}
in the magnetic arms [$W$ inherits the spin structure of Eq.\ \eqref{spinAll}, The action Eq.\ \eqref{SWF} is valid as long as $\nu LT\gg1$]. In the boundary action, we only keep terms linear in fluctuations on either side:
\begin{align}
S_{ij} = -\frac{g_t}{8}\tr\left[\mathcal{Q}_iW_ie^{\pi i\tau_3\frac{\hat\Phi_{ij}}{\Phi_0}} \mathcal{Q}_jW_je^{-\pi i\tau_3\frac{\hat\Phi_{ij}}{\Phi_0}} \right].
\end{align}
To calculate $\langle I_1I_2\rangle$ to the lowest order in $G_t$, we expand the exponent with respect to the boundary terms $S_{ij}$.
The leading terms with non-zero $\partial_{\varphi_1}\partial_{\varphi_2}$ contributing to Eq.\ \eqref{IQ} are as follows
\begin{multline}
\int e^{-\sum S_i-\sum S_{ij}}DW \mapsto  \langle S_{la}S_{ar}S_{rb}S_{bl}\rangle\\+
\frac14\langle S_{la}^2S_{ar}^2\rangle+\frac14\langle S_{rb}^2S_{bl}^2\rangle. \label{W8}
\end{multline}
The averages are with respect to the Gaussian bulk action (\ref{SWS})-(\ref{SWF}).

In the H case, only the first term in the right-hand side of Eq.\ \eqref{W8} is nonzero. Rewriting this term explicitly, we get
\begin{multline}
\left\langle
\tr\left[\mathcal{Q}_lW_le^{\pi i\tau_3\frac{\hat\Phi_{la}}{\Phi_0}} \mathcal{Q}_aW_a(0)e^{-\pi i\tau_3\frac{\hat\Phi_{la}}{\Phi_0}} \right]\right.\\
\times \tr\left[\mathcal{Q}_aW_a(L)e^{\pi i\tau_3\frac{\hat\Phi_{ar}}{\Phi_0}} \mathcal{Q}_rW_re^{-\pi i\tau_3\frac{\hat\Phi_{ar}}{\Phi_0}} \right]\\
\times \tr\left[\mathcal{Q}_rW_re^{\pi i\tau_3\frac{\hat\Phi_{rb}}{\Phi_0}} \mathcal{Q}_bW_b(L)e^{-\pi i\tau_3\frac{\hat\Phi_{rb}}{\Phi_0}} \right]\\
\left. \times \tr\left[\mathcal{Q}_bW_b(0)e^{\pi i\tau_3\frac{\hat\Phi_{bl}}{\Phi_0}} \mathcal{Q}_lW_le^{-\pi i\tau_3\frac{\hat\Phi_{bl}}{\Phi_0}} \right]
\right\rangle. \label{eq1234}
\end{multline}
Here, $W_a(0)$ and $W_a(L)$ stand for the fluctuation $W$ in arm $a$ at its contact with the left and right superconducting lead, respectively. $W_l$ and $W_r$ are the fluctuations on the superconducting side of that contact (left and right, respectively). For the superconducting fluctuations we omitted the coordinate, since the contacts are assumed to be close to each other, within a distance $d\ll\xi$. Fluctuations in the four different regions (two superconductors and two arms) are uncorrelated, so that the averaging can be done independently in each of them. The correlator between matrix elements of $W$ at the opposite ends of an arm follows from Eq.\ \eqref{SWF} and the symmetries imposed on $W$:
\begin{multline}
\langle W^{ij}_a(0)W^{kl}_a(L)\rangle \\ = \frac{\sum_u \bigl[ u^{il} u^{kj} - (C u)^{jl} (u C^T)^{ki} \bigr]}
         {8 \pi \nu L\sqrt{E_{\mathrm{Th}} (|\omega_i| + |\omega_j|)} \sinh \sqrt{(|\omega_i| + |\omega_j|)/E_\mathrm{Th}}}. \label{eq_diffuson}
\end{multline}
Here, $E_\mathrm{Th}=\hbar D/L^2$ is the Thouless energy of an arm. The summation over $u$ runs over the set of 16 matrices $ \{ 1, i\mathcal{Q} \} \otimes \{ 1, \tau_3 \} \otimes \{ 1, s_1, s_2, s_3 \}$, ensuring the symmetries $W\mathcal{Q}+\mathcal{Q}W = [W,\tau_3] = [W, \mathbf{s}] = 0$.

Expression \eqref{eq_diffuson} is the diffuson propagator from $0$ to $L$, it describes the diffusion of an electron-hole pair across the magnetic arm.

For the superconducting parts we have a same-point fluctuation correlator, derived from Eq.\ \eqref{SWS}:
\begin{equation}
 \langle W^{ij} W^{kl} \rangle =
 \frac{\sum_v \bigl[ v^{il} v^{kj} - (C v)^{jl} (v C^T)^{ki} \bigr]}
         {\pi \nu_s \sqrt{\hbar D_s \bigl( \sqrt{\omega_i^2 + \Delta^2} + \sqrt{\omega_{\smash{j}}^2 + \Delta^2} \bigr)}},
 \label{eq_diffusonSC}
\end{equation}
with $v \in \{1,i\mathcal{Q}\}$.

The product Eq.\ \eqref{eq1234} corresponds to the diagram in Fig.~\ref{fig1} (right). Two diffusons join the superconductors through the arms $a$ and $b$. They are then connected inside the superconductors recombining into $s$-wave Cooper pairs. The two other terms in Eq.\ \eqref{W8} (the second and the third one) are nonzero in the M case and generate diffusons travelling through the same arm, as shown in Fig.~\ref{fig2}. They are calculated similarly to Eq.\ \eqref{eq1234}.
\section{Results}
Plugging the correlators (\ref{eq_diffuson}) and (\ref{eq_diffusonSC}) into Eq.\ \eqref{eq1234}, we calculate $\langle I_1I_2\rangle$. The result is a function of $\delta\varphi=\varphi_1-\varphi_2$ and $\delta\Phi=\Phi_1-\Phi_2$:
\begin{subnumcases}{\label{IIAll}
\langle I_1 I_2\rangle = \frac{I_0^2}2 \cos(\delta\varphi)}
\sin^4\frac{\alpha}{2}, &H,\qquad\label{IIH}\\
\frac12\cos^2\left[\frac{\pi\delta\Phi}{2\Phi_0}\right], &M,\label{IIM}
\end{subnumcases}
where $\alpha$ is the angle between polarizations $\mathbf{n}_a$ and $\mathbf{n}_b$ of the arms in the H case, and
\begin{multline}
I_0 = \frac{4\sqrt{2}e E_{\mathrm{Th}}}{\hbar}\frac{G_t^2}{G_LG_\xi}\biggl\{T^2\sum\limits_{\omega_{1,2}>0}\biggl[\left(1+\frac{\omega_1^2}{\Delta^2}\right)\left(1+\frac{\omega_2^2}{\Delta^2}\right) \\ \times\left(\sqrt{1+\omega_1^2/\Delta^2}+\sqrt{1+\omega_2^2/\Delta^2}\right)\\ \times(\omega_1+\omega_2)E_\mathrm{Th}\sinh^2\sqrt{(\omega_1+\omega_2)/E_{\mathrm{Th}}} \biggr]^{-1}\biggr\}^{1/2}, \label{mag}
\end{multline}
gives the typical current magnitude in the SQUIDs. Here $G_L$ is the conductance of an arm, and $G_\xi = 2e^2\nu_s D_s/\xi$ is the normal-state conductance of the leads on length $\xi=\sqrt{\hbar D_s/\Delta}$. In the limit $E_\mathrm{Th}\ll\Delta$, Eq.\ \eqref{mag} reproduces the expression for mesoscopic supercurrent fluctuations in a single long SFS junction \cite{Zyuzin2003}.

Supercurrent fluctuations in an SFS SQUID have been studied before for the case $\alpha=\pi$, $\Phi=0$ in the absence of orbital magnetic effects \cite{Melin2005},
however, the result was different.
Expressions of Ref.~\cite{Melin2005} in the limit of zero temperature and infinite phase coherence length (this parameter is assumed to be infinite in our calculation from the beginning) produce the fluctuational current that does not depend on the length of the magnetic arms. On the contrary, our result (\ref{mag}) does depend on $L$ in the same limit. The suppression of the supercurrent with $L$, as predicted by our Eq.\ (\ref{mag}), is physically expectable. In our opinion, the discrepancy of the results can be due to a technical problem \cite{commentMelin} in the derivation of Ref.~\cite{Melin2005}.

In the most interesting limit of low temperature, $T\ll E_\mathrm{Th},\Delta$, Eq.\ \eqref{mag} simplifies to
\begin{align}
I_0 = \frac{4eE_\mathrm{Th}}{\sqrt{\pi}\hbar}\frac{G_t^2}{G_LG_\xi}\sqrt{\ln\left[\frac{\min(\Delta,E_{\mathrm{Th}})}{T}\right]}. \label{IlowT}
\end{align}
Equations (\ref{mag}) and (\ref{IlowT}) are applicable as long as $G_t^2E_\mathrm{Th}\ll G_LG_\xi T$.

In the above calculations we assumed the leads and the arms to be much narrower than $\xi$. This meant that transverse fluctuations could be neglected, so that $Q(x)$ was a one-dimensional field. We also assumed the arms to contact a superconductor within a distance $d\ll\xi$ from each other, so that the one-dimensional correlator \eqref{eq_diffusonSC} is taken at coincident points.

Our calculation is easily adapted to the geometry of three-dimensional superconductor leads. We still assume that the magnetic arms have a small radius $r\ll\xi$, but the distance $d$ between the contacts can now be large. The only change to our calculation scheme for the split-pair current is replacing Eq.\ \eqref{eq_diffusonSC} with a three-dimensional correlator $\langle W^{ij}(0)W^{kl}(d)\rangle$. The new correlator retains the symmetries of Eq.\ \eqref{eq_diffusonSC}, but now depends on $d$. The result can be obtained from Eq.\ \eqref{IlowT} by replacing
\begin{equation}
\frac{1}{G_\xi}\mapsto\frac{4\pi e^{-d/\xi}}{e^2\nu_{3d}D_sd}. \label{eq3d}
\end{equation}
The small-$d$ divergence of Eq.\ \eqref{eq3d} is cut off at $d\sim r$. At large $d/\xi$, the split-pair current is exponentially suppressed. This shows that the arms must be connected to each lead within the distance of the order of $\xi$ from each other for the Cooper pair splitting to be viable \cite{Melin2005}. Indeed, the size of a Cooper pair is $\sim\xi$. If the two contacts are much further apart than $\xi$, splitting into the two arms becomes a tunneling process corresponding to an unpaired electron traveling a distance of $d$ through the gapped superconductor. At the same time, obviously, same-arm currents do not depend on $d$.

We conclude that the magnitude of the currents we study is maximal when $d\lesssim\xi$. This should not be too restrictive experimentally, since dirty superconductors typically show $\xi$ of order of $10..100$ nm.

\begin{figure}
\centering
\hspace*{-3pt}\includegraphics[width=0.400\textwidth]{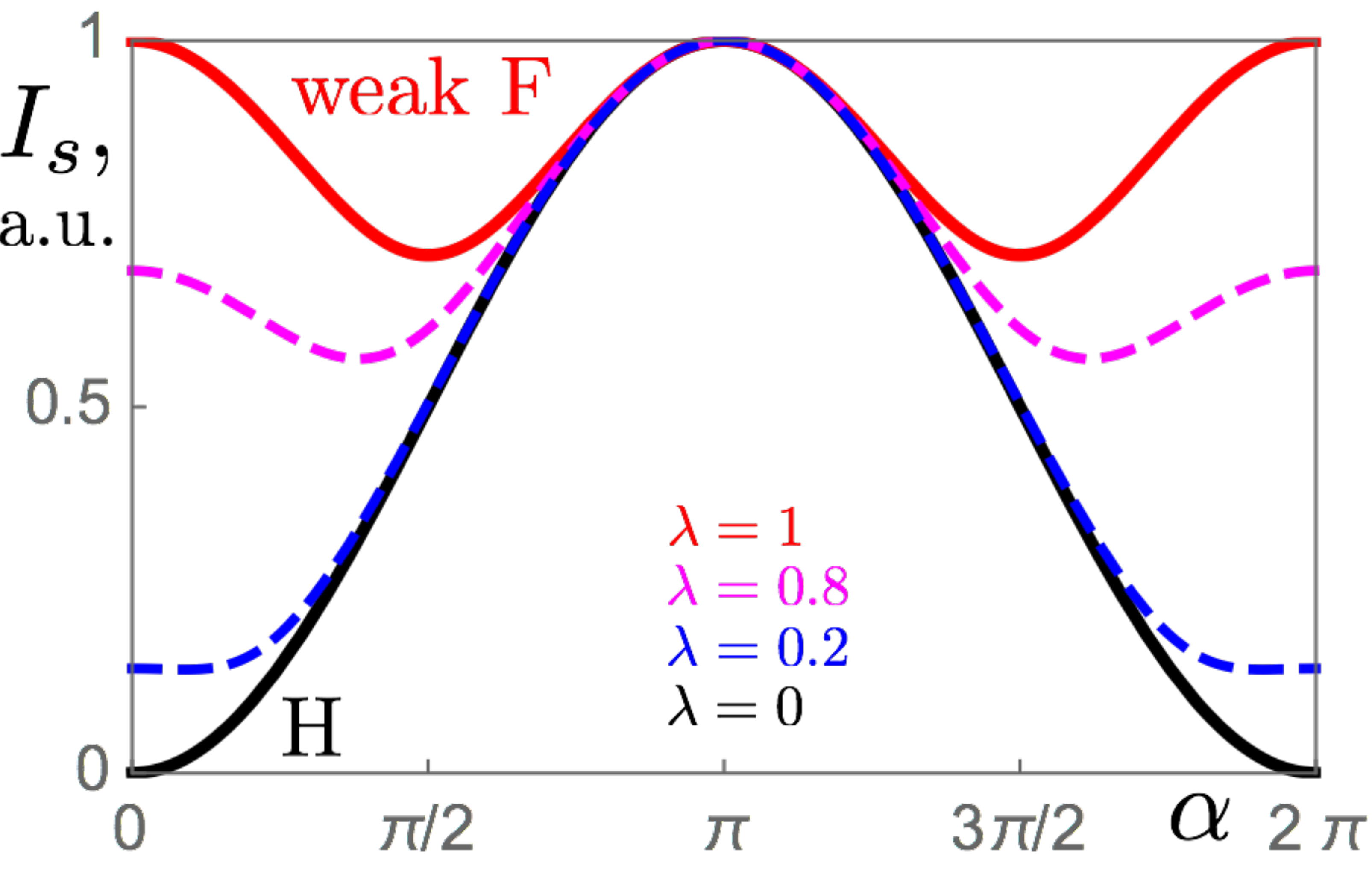}
\caption{
Angular dependence of the split-pair current for different degrees of polarization, Eq.\ \eqref{IF}. The bottom curve describes the H case where $I_s\propto \sin^2(\alpha/2)$. The top curve describes the limit of a weakly polarized ferromagnet where $I_s\propto \sqrt{1+\cos^2\alpha}$. Dashed curves describe intermediate polarization strengths.
}
\label{fig3}
\end{figure}
\subsection{Sample-specific current}
In Eqs.\ (\ref{IIAll})-(\ref{IlowT}), we calculated the current-current correlator, and obtained the typical magnitude of the fluctuational current. We can also extract the phase and flux dependence $I(\varphi,\Phi)$ of specific samples (i.e., for a specific disorder realization) from Eq.\ \eqref{IIAll}. Owing to the tunneling limit, only the first harmonic enters the current \cite{commentCurrent}.
In the H case, from Eq.\ \eqref{IIH} we get
\begin{subequations}\label{IAll}
\begin{align}
I(\varphi,\Phi) = I_{ab}\sin^2\frac{\alpha}2\sin(\varphi-\varphi_{ab}). \label{IH}
\end{align}
The amplitude $I_{ab}$ and phase $\varphi_{ab}$ are sample-specific constants. $I_{ab}$ has the root mean square (r.m.s.) $\sqrt{\langle I_{ab}^2\rangle}=I_0$, and $\varphi_{ab}$ is uniformly distributed. Note that Eq.\ \eqref{IH} does not contain $\Phi$ -- the current through the SHS SQUID is insensitive to the flux. This is because the current is carried exclusively by split Cooper pairs. In our gauge, the electron travelling through $a$ picks up a vector-potential-related phase $\varphi/2+\pi\Phi/2\Phi_0$, while its partner picks up $\varphi/2-\pi\Phi/2\Phi_0$. Thus, the total phase felt by the pair is $\varphi$, and the flux drops out.

In the M case, the current is obtained from Eq.\ \eqref{IIM},
\begin{multline}
I(\varphi,\Phi) = \frac{I_{ab}}2\sin(\varphi-\varphi_{ab})\\+
\frac{I_a}{2^{\frac32}}\sin\left(\varphi-\varphi_a+\pi\frac{\Phi}{\Phi_0}\right)+\frac{I_b}{2^{\frac32}}\sin\left(\varphi-\varphi_b-\pi\frac{\Phi}{\Phi_0}\right).\label{IM}
\end{multline}
The three terms correspond to the three diagrams in Fig.~\ref{fig2}, which sum up to Eq.\ \eqref{IIM}. Physically, the three terms in Eq.\ \eqref{IM} correspond to three transport processes. The first describes Cooper pairs split between the arms, just as in the SHS system. The other two describe the current from Cooper pairs that travel through a single arm. The sample-specific amplitudes $I_{ab}$, $I_a$, $I_b$ and phases $\varphi_{ab}$, $\varphi_a$, $\varphi_b$ are all uncorrelated. Phases are distributed uniformly, while the r.m.s.\ of $I_{ab}$, $I_a$, $I_b$ all equal $I_0$.

Finally, let us consider the F case, i.e., ferromagnetic arms that are not fully polarized. Each such arm can be viewed as a superposition of two half-metal arms with opposite polarizations. The SQUID then effectively contains four half-metallic channels, each supporting a single polarized diffuson mode \cite{commentF}. Consequently $\langle I_1I_2\rangle$ consists of six diagrams, each describing two diffusons travelling through a particular pair of half-metal channels (four diagrams for split-pair transport, and two for same-arm transport). We calculate these diagrams, taking into account different densities of states $\nu_+$ and $\nu_-$ for differently polarized electrons in the arms \cite{commentSpin}, and extract $I(\varphi,\Phi,\alpha)$:
\begin{multline}
I(\varphi,\Phi) = I_{ab}\sin(\varphi-\varphi_{ab})\sqrt{\sin^4\frac\alpha2+\frac{\lambda^2}{2}
\cos\alpha}
\\
+\frac{\lambda I_a}{2}\sin\left(\varphi-\varphi_a+\pi\frac{\Phi}{\Phi_0}\right)+\frac{\lambda I_b}{2}\sin\left(\varphi-\varphi_b-\pi\frac{\Phi}{\Phi_0}\right).\label{IF}
\end{multline}
where $\lambda=2\sqrt{\nu_+\nu_-}/(\nu_++\nu_-)$ characterizes the degree of spin polarization. It changes from $\lambda=0$ in the H case [in which case Eq.\ \eqref{IH} is reproduced] to $\lambda\to1$ in the weak ferromagnet case. The $\alpha$ dependence of the split-pair current $I_s(\alpha)$ (the $ab$ term) is shown in Fig.~\ref{fig3}. While in the H case $I_s(\alpha)$ is absent for $\alpha=0$ and grows with $\alpha$ to reach a maximum at $\pi$, the weak ferromagnetic limit $\lambda\to1$ is very different. Indeed, $I_s(\alpha)$ shows maxima at $\pi k$ and minima at $\pi (k+1/2)$. The $\pi$ periodicity in this limit is natural. Indeed, when $\nu_+=\nu_-$, the pair of effective H channels constituting the F arm is the same for polarizations $\mathbf{n}$ and $-\mathbf{n}$.

\end{subequations}

\section{Discussion}
Equations (\ref{mag})-(\ref{IAll}) are our central results and provide a clear physical picture of the Josephson effect in magnetic SQUIDS where it is dominated by mesoscopic fluctuations. The current in these systems is of the same order for
H, F, and M arms. However, the dependence on polarization orientation $\alpha$ and flux $\Phi$ is quite different in the three cases.

In a conventional SNS SQUID, the fluctuational contribution due to split Cooper pairs exists as well, however, the coherent contributions of each arm are much larger (and immune to impurity averaging) \cite{commentSNS}. In our case, the current is all fluctuational, and comparable contributions come from all the three processes -- pairs going through $a$, pairs going through $b$, and pairs splitting between the arms. The presence of the latter process leads to the $2\Phi_0$ periodicity of the critical current, see Eqs.\ (\ref{IM})-(\ref{IF}), in contrast to conventional $\Phi_0$-periodic dependence.
This effect is similar to the one discussed in Refs.~\cite{SplitterQD}.

The randomness of the phase shifts in Eqs.\ (\ref{IAll}) means that the junctions are $\varphi_0$ junctions in the sense that the current-phase relation (for a specific disorder realization) can be represented in the form $I(\varphi)= I_c \sin(\varphi-\varphi_0)$. Such junctions, which can be used as phase batteries, are now actively studied both theoretically and experimentally. The $\varphi_0$ phase offset only emerges if TRS is fully broken. For example, this can happen due to the simultaneous presence of the Zeeman field and spin-orbit interaction \cite{Buzdin2008,Kouwenhoven2016,Mironov2015} or due to several noncoplanar directions of the Zeeman field \cite{Liu2010,commentTRS}. On the contrary, if the weak link of a Josephson junction involves coplanar exchange fields, $\varphi_0$ is restricted to $0$ or $\pi$ only \cite{BuzdinReview2005,BVE,commentTRS}.

The latter result was formulated for the average current, which is exponentially suppressed in our case. At the same time, our results indicate that the fluctuational current can demonstrate $\varphi_0$-junction behavior even if the averaged system only allows $\varphi_0=0,\pi$. Moreover, the equilibrium phase $\varphi_0$ in our system can be controlled by the external flux $\Phi$ in the SMS and SFS cases, Eqs.\ (\ref{IM}) and (\ref{IF}).

Let us now discuss the amplitude of the current, Eq.\ \eqref{IlowT}. The current is fluctuational and hence small. Compared to an SNS system of the same geometry (i.e., same sizes and normal transport characteristics, but with nonmagnetic metallic arms), the current is suppressed by the small factor $e^2/\hbar G_\xi$, up to logarithmic factors. Nevertheless, the current in our magnetic SQUIDs can be large enough to be detected experimentally. We take $G_t^2/G_LG_\xi\sim1$ and $T\sim E_\mathrm{Th}$, which is on the border of applicability of our model, so that $I_0\sim eE_\mathrm{Th}/\hbar$. This can reach several nA in a system with $E_\mathrm{Th} \sim 1$~K.

Effects of the Cooper pair splitting in magnetic SQUIDs resemble those discussed in Refs.~\cite{SplitterQD} for quantum-dot setups.
While the $\alpha$ dependence of the Josephson current in Eq.\ (\ref{IH}) and $\Phi$ dependence in Eq.\ (\ref{IM}) are similar to Refs.~\cite{SplitterQD}, there are also essential differences. The disordered nature of our system and full TRS breaking lead to a vanishing average current, so that only fluctuations survive, and the fluctuational split (and total) current demonstrates $\varphi_0$-junction behavior. In addition, we have shown the possibility of magnetization-direction control of the ratio between conventional SQUID transport and split-pair current, see Eq.\ \eqref{IF}.

\section{Conclusions}
To conclude, we have calculated the Josephson current carried by split Cooper pairs in an SHS SQUID, as well as its SMS and SFS counterparts. In the former case, the current is carried exclusively by split Cooper pairs, depends on the relative magnetization direction of the arms and is insensitive to flux. In the SFS and SMS systems, the current depends on the flux with a period $2\Phi_0$, which is twice the period of a conventional SQUID. In all the cases, the junctions turn out to be in the $\varphi_0$ state, which is adjustable by magnetic flux in SMS and SFS junctions. The sample-specific supercurrent in all studied systems is produced by mesoscopic fluctuations. Consequently, the current is suppressed by a factor of $e^2/\hbar G_\xi$ compared to an SNS device of the same geometry ($G_\xi$ is the normal-state conductance of the leads on the length $\xi$). The current is nevertheless large enough to be observed experimentally, and can reach several nA for a properly designed system.

\acknowledgments
The idea of this research was formulated in the course of our conversations with V.~V.\ Ryazanov. We are also grateful to him for useful discussions of our results. The field-theoretical calculation was supported by the Russian Science Foundation (Grant No.\ 14-12-00898).
Ya.F. was supported in part by the RF Ministry of Education and Science (Grant No.\ 14Y.26.31.0007).
The research was also partially supported by the RF Presidential Grant No.\ NSh-10129.2016.2.


\begin{thebibliography}{99}

\bibitem{EPR}
A.\ Einstein, B.\ Podolsky, and. N.\ Rosen,
Phys.\ Rev.\ \textbf{47}, 777 (1935).

\bibitem{Bell}
J.~S.\ Bell, Physics (Long Island City, NY) \textbf{1}, 195 (1965);
J.~S.\ Bell, Rev.\ Mod.\ Phys.\ \textbf{38}, 447 (1966);
J.~F.\ Clauser, M.~A.\ Horne, A.\ Shimony, and R.~A.\ Holt, Phys.\ Rev.\ Lett.\ \textbf{23}, 880 (1969).



\bibitem{Recher2001}
P.\ Recher, E.~V.\ Sukhorukov, and D.\ Loss,
Phys.\ Rev.~B \textbf{63}, 165314 (2001).

\bibitem{Lesovik2001}
G.~B.\ Lesovik, T.\ Martin, and G.\ Blatter,
Eur.\ Phys.~J.~B \textbf{24}, 287 (2001)

\bibitem{Borlin2002}
J.\ B\"{o}rlin, W.\ Belzig, and C.\ Bruder,
Phys.\ Rev.\ Lett.\ \textbf{88}, 197001 (2002).

\bibitem{Chtchelkatchev2002}
N.~M.\ Chtchelkatchev, G.\ Blatter, G.~B.\ Lesovik, and T.\ Martin,
Phys.\ Rev.~B \textbf{66}, 161320(R) (2002).

\bibitem{Hofstetter2009}
L.\ Hofstetter, S.\ Csonka, J.\ Nyg\aa rd, and C.\ Sch\"{o}nenberger,
Nature \textbf{461}, 960 (2009).

\bibitem{Herrmann2010}
L.~G.\ Herrmann, F.\ Portier, P.\ Roche, A.~L.\ Yeyati, T.\ Kontos, C.\ Strunk,
Phys.\ Rev.\ Lett.\ \textbf{104}, 026801 (2010).

\bibitem{Schindele2012}
J.\ Schindele, A.\ Baumgartner, C.\ Sch\"{o}nenberger,
Phys.\ Rev.\ Lett.\ \textbf{109}, 157002 (2012).

\bibitem{Das2012}
A.\ Das, Y.\ Ronen, M.\ Heiblum, D.\ Mahalu, A.~V.\ Kretinin, H.\ Shtrikman,
Nature Commun.\ \textbf{3}, 1165 (2012).

\bibitem{SplitterQD}
M.~S.\ Choi, C.\ Bruder, and D.\ Loss, Phys.\ Rev.~B \textbf{62}, 13569 (2000);
Z.\ Wang and X.\ Hu, Phys.\ Rev.\ Lett.\ \textbf{106}, 037002 (2011);
R.\ Jacquet, J.\ Rech, T.\ Jonckheere, A.\ Zazunov, and T.\ Martin, Phys.\ Rev.~B \textbf{92}, 235429 (2015).

\bibitem{Keizer2006}
R.\ Keizer, S.~T.~B.\ G\"onnenwein, T.~M.\ Klapwijk, G.\ Miao, G.\ Xiao, and A.\ Gupta,
Nature (London) \textbf{439}, 825 (2006).

\bibitem{Singh2015}
A.\ Singh, S.\ Voltan, K.\ Lahabi, and J.\ Aarts,
\textbf{5}, 021019 (2015).

\bibitem{Leksin2012} 
P.~V.\ Leksin, N.~N.\ Garif’yanov, I.~A.\ Garifullin, Ya.~V.\ Fominov, J.\ Schumann, Y.\ Krupskaya, V.\ Kataev, O.~G.\ Schmidt, and B.\ B\"uchner,
Phys.\ Rev.\ Lett.\ \textbf{109}, 057005 (2012).

\bibitem{BWL2004}
D.\ Beckmann, H.~B.\ Weber, and H.\ v.\ L\"ohneysen, Phys.\ Rev.\ Lett.\ \textbf{93}, 197003 (2004).

\bibitem{BL2005}
D.\ Beckmann and H.\ v.\ L\"ohneysen, AIP Conf.\ Proc.\ \textbf{850}, 875 (2006).

\bibitem{Zyuzin2003}
A.~Yu.\ Zyuzin, B.\ Spivak, and M.\ Hru\v{s}ka,
Europhys.\ Lett.\ \textbf{62}, 97 (2003).

\bibitem{Melin2003}
R.\ M\'elin and S.\ Peysson,
Phys.\ Rev.~B \textbf{68}, 174515 (2003).

\bibitem{Wegnersigma}
F.\ Wegner, Z.~Phys.~B \textbf{35}, 207 (1979); L.\ Sch\"afer and F.\ Wegner, Z.~Phys.~B \textbf{38}, 113 (1980).

\bibitem{Altlandsigma}
A.\ Altland, B.~D.\ Simons, and D.\ Taras-Semchuk, Adv.\ Phys.\ \textbf{49}, 321 (2000).

\bibitem{Efetov}
K.~B.\ Efetov, \textit{Supersymmetry in Disorder and Chaos} (Cambridge Univ.\ Press, Cambridge, England, 1996).

\bibitem{Usadel1970}
K.~D.\ Usadel, Phys.\ Rev.\ Lett.\ \textbf{25}, 507 (1970).

\bibitem{Melin2005}
R.\ M\'elin,
Phys.\ Rev.~B \textbf{72}, 134508 (2005).



\bibitem{commentMelin}
Our Eq.\ \eqref{IQ} for the current-current correlator contains two frequency summations (coming from expressions for $I_1$ and $I_2$). In contrast to that, Eq.\ (14) in Ref.\ \cite{Melin2005} contains only one energy integration (in the Keldysh technique, it substitutes the frequency summation). We doubt the validity of such representation. Note that two energy integrations (or summations) gather spectral contributions to each current from quasiparticles of different energies, and there are no reasons for reduction to a single integration.






\bibitem{commentCurrent}
This can also be derived directly from Eq.\ \eqref{IIAll}. E.g., for the H case, we rewrite the phase dependence of $\langle I_1I_2\rangle$ as $\cos\varphi_1\cos\varphi_2+\sin\varphi_1\sin\varphi_2$. Expanding $I(\varphi)$ in the basis of $\{\cos n\varphi,\ \sin n\varphi\}$, we see that only the first harmonic is allowed by the correlator. The M case is treated similarly.

\bibitem{commentF}
Using the same $\sigma$-model method, one can show rigorously that this trick is valid. The $Q$ matrix in a ferromagnet obeys $[Q,\mathbf{n}\boldsymbol{\sigma}]=0$, which uncouples the diffusion of electrons with spin along or against $\mathbf{n}$.




\bibitem{commentSpin}
We also assume  spin-dependent tunnelling conductance, $G_{t}\propto \nu_\pm$.

\bibitem{commentSNS}
Fluctuational contributions from same-arm processes in SNS junctions are also small (compared to the average same-arm current)  \cite{SkvortsovHouzet}.

\bibitem{SkvortsovHouzet}
M.\ Houzet and M.\ A.\ Skvortsov,
Phys.\ Rev.\ B \textbf{77}, 024525 (2008).

\bibitem{Buzdin2008}
A.\ Buzdin,
Phys.\ Rev.\ Lett.\ \textbf{101}, 107005 (2008).

\bibitem{Kouwenhoven2016}
D.~B.\ Szombati, S.\ Nadj-Perge, D.\ Car, S.~R.\ Plissard, E.~P.~A.~M.\ Bakkers, and L.~P.\ Kouwenhoven,
Nature Phys.\ \textbf{12}, 568 (2016).

\bibitem{Mironov2015}
S.~V.\ Mironov, A.~S.\ Mel'nikov, and A.~I.\ Buzdin, Phys.\ Rev.\ Lett.\ \textbf{114}, 227001 (2015).

\bibitem{Liu2010}
J.-F.\ Liu and K.~S.\ Chan, Phys.\ Rev.~B \textbf{82}, 184533 (2010).

\bibitem{commentTRS}
Note that coplanar exchange fields still respect TRS in the following sense: $H_Z=H_Z^*$ in a basis where the exchange fields are in the $xz$ plane, so that the Zeeman part of the Hamiltonian is $H_Z=h_x(\mathbf{r})\sigma_x+h_z(\mathbf{r})\sigma_z$.

\bibitem{BuzdinReview2005}
A.~I.\ Buzdin,
Rev.\ Mod.\ Phys.\ \textbf{77}, 935 (2005).

\bibitem{BVE}
F.\ S.\ Bergeret, A.\ F.\ Volkov, and K.\ B.\ Efetov,
Rev.\ Mod.\ Phys.\ \textbf{77}, 1321 (2005).




\end{thebibliography}
\end{document}